\documentclass[article]{aastex}

\def\percc{\hbox{cm$^{-3}$}}
\def\percmsq{\hbox{cm$^{-2}$}}

\def\TKIN{\hbox{T$_{\rm kin}$}}
\def\H2*{\hbox{H$_2^*$}}

\def\HII {\hbox{H~{\sc ii }}}
\def\H#1#2{\hbox{H~#1#2$\alpha$}}

\def\kms{~km~s$^{-1}$~}
\def\kmsns{~km~s$^{-1}$}

\def\eg{e.g.,~}

\def\etal{{et~al.~}}

\def\NH3{\hbox{NH$_3$}}
\def\INH3{\hbox{$^{15}$NH$_3$}}
\def\NINH3{\hbox{$^{14}$NH$_3$}}
\def\vlsr{\hbox{${\rm v}_{\rm LSR}$}}
\def\THCO{\hbox{$^{13}$CO }}
\def\CEIO{\hbox{C$^{18}$O}}

\def\MOLH{\hbox{H$_2$}}

\def\h2co{\hbox{H$_2$CO}}

\chardef\isp="10 \def\i{\'\isp}

\lefthead{}
\righthead{}

\begin{document}

\title{
Submillimeter CO Line Emission from Orion }
\author{ T. L. Wilson\altaffilmark{1}$^,$\altaffilmark{2},
D. Muders\altaffilmark{1}$^,$\altaffilmark{2}, C. Kramer\altaffilmark{3},
C. Henkel\altaffilmark{2}}

\altaffiltext{1}{Sub-Millimeter Telescope Observatory, Steward Observatory,
The University of Arizona, Tucson, Az., 85721}
\altaffiltext{2}{Max Planck Institut f\"{u}r Radioastronomie, Postfach 2024,
D-53010 Bonn, Germany}
\altaffiltext{3}{I. Physikalisches Inst., Univ. zu K\"oln, Z\"ulpicherstr.
77, D-50937 K\"oln, Germany}

\begin{abstract}
Images of an 8 square minute region around the Orion KL source have been 
made in the $J=7-6$ (806 GHz) and $J=4-3$ (461 GHz) lines of CO with 
angular resolutions of 13$''$ and 18$''$. These data were taken 
employing on-the-fly mapping and position switching techniques. Our 
$J=7-6$ data set is the largest image of Orion with the highest 
sensitivity and resolution obtained so far in this line. Most of the 
extended emission arises from a Photon Dominated Region (PDR), but 8\% 
is associated with the Orion ridge. For the prominent Orion KL outflow, 
we produced ratios of the integrated intensities of our $J=7-6$ and $4-
3$ data to the $J=2-1$ line of CO. Large Velocity Gradient (LVG) models 
fit the outflow ratios better than PDR models. The LVG models give H$_2$ 
densities of $\sim$10$^5$ \percc. The CO outflow is probably heated by 
shocks.  In the Orion~S outflow, the CO line intensities are lower than 
for Orion~KL. The $4-3$/$2-1$ line ratio is 1.3 for the blue shifted 
wing and 0.8 for the red shifted wing. Emission in the jet feature 
extending 2$'$ to the SW of Orion~S was detected in the $J=4-3$ but not 
the $J=7-6$ line; the average $4-3$/$2-1$ line ratio is $\sim$1. The 
line ratios in the Orion~S outflow and jet features are consistent with both 
PDR and LVG models.Comparisons of the intensities of the $J=7-6$ and $J=4-3$ 
lines from the Orion Bar with PDR models show that the ratios exceed 
predictions by a factor of 2. Either clumping or additional heating by 
mechanisms, such as shocks, may be the cause of this discrepancy. 
\end{abstract} 

\section{Introduction}

The OMC-1 region is the closest molecular cloud where high mass O-B star 
formation has recently taken place (see, \eg the review of O'Dell 2001). 
The region within 3$'$ of Orion KL is a particularly fruitful object of 
study. There is a chemically rich, dense warm region, the `Hot Core' 
(see Wilson \etal 2000 and references therein), two outflow sources 
(see, e.g., Rodr{\i}guez-Franco, Mart{\i}n-Pintado \& Wilson 1999; 
hereafter RMW, Gaume et al.~1998, McMullin, Mundy \& Blake 1993) and 
extended warm gas from a PDR at the interface with the rear boundary of 
the Orion \HII region. Behind the PDR is the `Orion ridge'. This is part 
of the column-like feature extending north-south over 2$^{\rm o}$ (see, 
\eg Tatematsu \etal 1993). Near Orion KL, there is a rapid change in 
radial velocity in the ridge. This is caused by the presence of a number 
of separate clouds with different radial velocities (Womack, Ziurys \& 
Sage 1993; Wang, Wouterloot \& Wilson 1993). In addition, there is 
another neutral-ionized gas interface, the Orion Bar, SW of the \HII 
region (see van der Werf et al.~1996). 

Spatially extended emission from warm molecular and atomic gas arises in 
PDR's. In PDR's, the kinetic temperatures reach hundreds of  degrees 
(see Hollenbach \& Tielens 1999). Thus the $J=7-6$ line of CO, emitted 
from an energy level 156 K above ground, should be a good tracer of 
molecular gas PDR's. There is a partial map of this region in the $J=7-
6$ line by Howe et al.~(1993, 20$''$ beam) and a complete map by Schmid-
Burgk et al.~(1989, 98$''$ beam). Schmid-Burgk \etal (1989) used beam 
switching with chopper throws of $<$6$'$. Since the CO is more extended, 
this chopping resulted in confusion between CO emission in the signal 
and reference beams. Schulz et al.~(1995; 15$''$ beam) also mapped the 
Orion KL region in the $J=4-3$ line of CO, but with a telescope which 
had a low beam efficiency and from a site closer to sea level. In order 
to trace the extent of warm gas, to compare CO emission from high and 
low $J$ rotational levels, and to relate molecular emission to compact 
continuum sources (see Mezger, Zylka \& Wink 1990, Menten \& Reid 1995), 
we have made position-switched images of the Orion KL region in the 
$J=4-3$ and $J=7-6$ lines. 

    \section{ Observations}

The $J=4-3$ and $J=7-6$ line CO data were taken with the 10-meter
diameter Heinrich Hertz Telescope (HHT) on Mt. Graham, AZ {\footnote {The 
HHT  is operated by the Submillimeter Telescope Observatory on behalf of 
the Max-Planck-Institut f. Radioastronomie and Steward Observatory of 
the University of Arizona.}}. At the $J=4-3$ line, at 461.041 GHz, the 
FWHP beam size is 18$''$. During the $J=4-3$ and $J=7-6$ line observing 
sessions, the pointing accuracy, from measurements of Saturn, was better 
than 2$''$. Since the pointing model for each receiver is merely a 
constant offset from a general pointing model, we are confident that the 
RMS pointing accuracy is 2.5$''$, the usual value (Wilson et al.~2001). 
The $J=4-3$ line data were taken on Feb.~3, 1999, Feb.~7, 2000 and 
Feb.~16, 2001 with a single channel SIS mixer facility instrument at the 
HHT. The single sideband receiver noise temperature was 260 K. During 
observations in Feb. 1999, the system noise including corrections for 
the atmosphere was $\sim$1200 K; in Feb. 2000 and 2001, the system noise 
was $\sim$3500 K. In 1999, we took spectra spaced by one full beam 
width; these data did not include the Orion Bar feature, and the jet 
feature SW of Orion~S, so in Feb.~2000, we mapped these regions using 
the On-The-Fly (OTF) technique. All of the OTF maps are fully sampled. 
The Feb.~2001 data consisted of longer integrations on the red- and 
blue-shifted maxima in the Orion~S outflow. Scans taken toward Orion~KL 
served to calibrate the $4-3$ data in all three sessions. Our 
calibrations were made on the assumption of equal response in the signal 
and image sidebands. This was checked by a comparison of our peak 
temperature for the Orion~KL, or (0$''$, 0$''$) position (Fig.~1(a)) 
with that of Schulz \etal (1995). Our peak is 20\% lower; given our 
better beam efficiency and pointing and better weather, we prefer our 
values. 

The chopper wheel calibration provides a corrected antenna temperature, 
T$^*_{\rm A}$, appropriate for a very extended source. At 460 GHz, 
measurements of the Moon give an efficiency of $\sim$0.75, while 
measurements of Saturn (diameter $\sim$20$''$) give a main-beam 
efficiency of $\sim$0.47; this latter value was used to estimate T$_{\rm 
MB}$ for compact CO emission, such as from outflow sources. The CO 
emission from the Orion ridge is very extended NS, but is of limited 
extent EW. Thus, the appropriate efficiency is that measured for a 
moderately extended object. For such moderately extended gas, we have no 
measurements of the HHT efficiencies, so have used the geometric mean of 
the two measured  efficiencies, 0.6, to estimate T$_{\rm MB}$ for 
quiescent $J=4-3$ CO emission. 

The $J=7-6$ line, at 806.652 GHz, was measured on Feb. 12, 1999. The
FWHP beam width of the HHT is estimated to be 13$''$. The pointing at 806 
GHz is related to the pointing model by a constant offset, so a check 
made by measuring the positions of Jupiter and Saturn during the 
observing run, assured us that the pointing is better than 3$''$. The 
data were taken employing the OTF mapping technique with the reference 
offset by $\Delta \alpha=-15'$, $\Delta \delta=0'$ from the (0$''$, 
0$''$) position. The $J=7-6$ line data were taken using Hot Electron 
Bolometer (HEB) receiver from the Harvard-Smithsonian Center for 
Astrophysics. The calibration was made using the chopper wheel 
technique. Tests in the laboratory and at the HHT with a single sideband 
filter showed no deviations from equal response in the signal and image 
sidebands. In addition, comparisons with the data of Howe et al.~(1993) 
show excellent agreement. Thus we conclude that the receiver sideband 
responses are equal. The single sideband receiver noise temperature at 
the frequency of the $J=7-6$ line was $\sim$1800 K. During the $J=7-6$ 
line measurements the average system noise, including corrections for 
the atmosphere, was $\sim$14000 K. 

The calibration of the CO $J=7-6$ line is also based on the chopper wheel 
method. At 806 GHz, measurements of the Moon gave an efficiency of 
$\sim$0.75. Measurements of Saturn (diameter $\sim$20$''$) and Mars 
(diameter $\le$14$''$) give a main-beam efficiency of $\sim$0.40; this 
value was used to estimate T$_{\rm MB}$ for compact CO emission, such as 
from outflow sources. For CO emission from the Orion ridge, we have used 
the geometric mean of the two measured  efficiencies, 0.54, to estimate 
T$_{\rm MB}$. 

The efficiencies obtained from measurements of the planets and Moon
indicate that there is some error beam contribution at 806 GHz. This 
will tend to smooth the spatial distribution of the CO emission. 
However, none of the extended CO $J=7-6$ emission is caused by an error 
beam smearing the intense emission from the Orion KL outflow, since the 
very extended CO emission features in our map have much smaller line 
widths and different radial velocities. 

The $J=4-3$ and $J=7-6$ spectra were analyzed using an Acoustic Optical 
Spectrometer (AOS) with 1 MHz frequency resolution. At the $J=4-3$ line 
rest frequency, the spectrometer resolution is 0.65 km s$^{-1}$. At the 
$J=7-6$ line rest frequency, this is 0.37 km s$^{-1}$. 

To compare our sub-mm data with mm results, we took $J=2-1$ CO spectra 
with the HHT in Feb.~2000. At the $2-1$ line frequency, the angular 
resolution is 33$''$ and the main-beam efficiency is 0.78. Additional 
unpublished \THCO\ $J=1-0$ and CO $J=2-1$ data were taken with the IRAM 
30-m telescope by R. Mauersberger. The angular resolution of the \THCO\ 
data was 21$''$; the forward and beam efficiencies were 0.91. For CO 
$J=2-1$ line data, the beam size was 13$''$, the forward efficiency was 
0.8, and the beam efficiency was 0.45. For the $J=2-1$ CO data for the 
Orion KL and Orion~S outflows, we have been given access to the data 
published by RMW. 

\section{Results}

Fig. 1 contains sample spectra. Fig.~2 shows a series of velocity maps 
and Plate 1 is a color-code brightness plot of T$_{\rm A}^*$ integrated 
from $-150$ \kms to 150 \kmsns. Our CO $J=4-3$ line for the Orion KL 
nebula is in Fig.~1(a); the corresponding  $J=7-6$ CO line is in (b). 

The extended CO $J=7-6$ emission has two remarkable properties, best 
seen in the color plate. First, emission extends over the entire map, 
that is, warm molecular gas is present over a $\ge$3$'$ region. Second, 
the maximum of the extended emission does {\it not} peak on the 
Trapezium, but close to the Orion Ridge. Line parameters for the 
extended CO emission near the Trapezium are in Fig.~3. 

The lowest temperature in our map is 10 K, at (100$''$, $-80''$). Near 
the position of the Trapezium stars, the peak T$_{\rm A}^*$ is 80 K. The 
peak T$_{\rm A}^*$ value of the quiescent CO is $\sim$100 K. This is 
within 20$''$ of the (0$''$, 0$''$) position. For extended CO $J=7-6$ 
emission, the conversion factor from T$_{\rm A}^*$ to T$_{\rm MB}$ is 
1.85. The Planck correction was used to convert a T$_{\rm MB}$ of 185 K 
to a T$_{\rm ex}$ of 204 K. For extended, optically thick, thermalized 
emission T$_{\rm ex}$ is equal to the kinetic temperature, \TKIN. Since 
the CO has small scale clumping (next section), \TKIN\ is actually 
larger than the peak CO temperature. We discuss CO emission at the 
Declination of the Trapezium in the next section. 

The well known, prominent Orion KL outflow is near our (0$''$, 0$''$) 
offset (see, \eg RMW). Contour maps of two velocities in the CO outflow 
toward Orion KL are shown in Fig.~4. We discuss Orion KL in Section 4.2. 

Another discrete source is S~6 or Orion~S (Batrla \etal 1983, BWB83). 
This source has the IAU designation `[BWB83] 6'. In maps of sub-mm dust 
continuum (see Mezger \etal 1990; hereafter MZW) and Far Infrared (FIR) 
fine structure lines [C~II] and [O~I] (Herrmann \etal 1997), the peak 
intensities of Orion KL and Orion~S are comparable, but in the $J=4-3$ 
and $7-6$ lines, Orion~S is more diffuse and much weaker. The CO 
emission from Orion~S is best seen in our color plate, where we show the 
locations of a compact outflow (mapped by RMW). The spectra in 
Fig.~1(c)-(d) were taken toward the blue shifted and red-shifted outflow 
maxima. We plot the locations of emission from different species in 
Fig.~5. In the color plate, we show the location of the $\ge2'$ 
($\ge$0.3 pc) long highly collimated jet extending SW (Schmid-Burgk 
\etal 1990). In Fig.~1(e)-(g) we show spectra from this feature. We 
discuss Orion~S in Section 4.3. 

The spectrum in Fig.~1(h) was taken toward the Orion Bar feature; this 
shows the fit of 2 velocity components; the component at 4 \kms\ is not 
related to the Orion Bar feature. In Fig.~2, at \vlsr=9.43 to 12.28 
\kms, the Bar is in the SE. In the color plate and Fig.~2 the Bar 
appears as two maxima. We show the emission from different species in 
Fig.~6 and discuss this feature in Section 4.4. 

\section{Discussion}
\subsection{ The Widespread Quiescent CO $J=7-6$ Emission }

Here we present results and analysis for the extended warm CO emission, 
concentrating on the Declination of the Trapezium. The CO emission is 
thought to arise in the interface between the \HII region and the 
molecular cloud, a PDR (see Hollenbach \& Tielens 1999). The most 
prolific sources of UV radiation is the star $\theta ^1$ C Orionis, the 
brightest star in the Trapezium. If only the Trapezium heated the CO, 
and the geometry were plane parallel, one would expect the warm CO 
emission to peak at the location of $\theta ^1$ C. From plate 1, this is 
definitely {\it not} the case. Stacey \etal (1993) proposed that  
$\theta ^1$ C is located in a cavity in the molecular cloud. This 
proposal would reduce the amount of emission from the PDR toward  
$\theta ^1$ C, but does {\it not} explain the east-west asymmetry in 
this emission. 

To make a quantitative relation between the CO $J=7-6$ emission and the 
Trapezium stars, we plot parameters of CO line profiles at the 
Declination of $\theta ^1$C, at $\Delta \delta=-53''$. Given our high 
angular resolution, this offset allows us to avoid emission from the 
discrete sources Orion KL and Orion~S. In Fig.~3, we show gaussian fit 
parameters for the CO profiles. Relative to the R.~A. of $\theta ^1$C, 
all of the CO maxima are at $\Delta \alpha=-33''$ to $-38''$, close to 
the location of the Orion ridge. The warmer ($J=7-6$) CO peaks 5$''$ 
east of the cooler ($J=2-1$) CO. The column density of \THCO\ peaks 
9$''$ west of the warm CO and 4$''$ west of the cooler CO. The warmer CO 
is found closer to $\theta ^1$C. The offset of the CO $J=7-6$ line 
relative to $J=2-1$ line is small, but given the pointing accuracy, is 
significant. Since the warmer CO is offset in the direction of the 
Trapezium, we conclude that some of the CO in the Orion ridge is heated 
by the Trapezium stars. 

We use the $J=7-6$ temperatures in Fig.~3 to estimate the proportion of 
warm molecular gas in the PDR and Orion ridge at this Declination. We 
drew a smooth curve connecting $\Delta \alpha=100''$ with $-200''$; the 
area above this is the emission from the Orion ridge. The ratio of areas 
in the Orion ridge to that between 400$''$ and $-300''$ is 0.08. On this 
basis, 92\% of the warm molecular gas arises in the PDR, the remainder 
in the Orion ridge. Since there is no counterpart of the Orion ridge in 
maps of ionized gas (see, \eg Wilson \etal 1997), we conclude that lower 
mass stars heat this gas. According to Hillenbrand \& Hartmann (1998) 
there are about 4500 M$_\odot$ in stars within 12$'$ of the Trapezium, 
so there would be sufficient stars to provide heating in this part of 
the Orion ridge. According to Kaufman, Hollenbach \& Tielens  (1998), 
embedded stars would be the most efficient heating sources. 

From PDR models and measured CO line intensities, we determine a range 
of H$_2$ densities. We assume that the region is illuminated from one 
side only.  From Model B (Fig.~10a) of K\"oster \etal (1994; KSSS) the 
best agreement with data is for a density of $\sim$10$^6$ \percc. This 
is true for a range of radiation fields. A more recent PDR model, by 
Kaufman \etal (1999; KWHL) also allows estimates of H$_2$ densities from 
the ratio of the CO $J=6-5$ to $J=1-0$ lines. From interpolation, we 
obtain a density of 10$^5$ \percc. We take this density as the best 
estimate. Toward the Trapezium, Rodr{\i}guez-Franco \etal (2001) found a 
density of 2 10$^6$ \percc\ from a multi-transition analysis of CN data. 

In PDR's, one expects strong emission from [C~II] and [O~I], so we have 
plotted the integrated line intensities (from Herrmann \etal 1997) in 
Fig.~3(d); the maps have angular resolutions of $\sim$1$'$, which leads 
to some confusion in a region as complex as Orion. However, the general 
morphology of the [C~II] and [O~I] images agree with our warm CO image 
in the color plate. We have applied the analysis of KWHL to the [C~II] 
and [O~I] fine structure line data of Herrmann \etal (1997) to obtain 
n(H$_2$)=3 10$^4$ \percc, which we take to be the average density. To 
reconcile the CO, [C~II] and [O~I] line data, the CO clumps must have a 
volume filling factor of 0.3. Since we can `see' the Orion ridge in the 
optically thick CO $J=7-6$ line the foreground CO in the PDR must be 
clumped. If the volume filling factor is 0.3, our peak $J=7-6$ Planck 
temperature, 204 K, becomes a \TKIN\ of 680 K. If H$_2$ densities in the 
CO emitting region are $> 10^5$ \percc, the volume filling factor will 
become smaller and \TKIN\ will rise. 

We can estimate the total distance from the PDR surface (also referred to 
as the `Main Ionization Front' by O'Dell 2001) to the Orion ridge, from 
our H$_2$ density of 3 10$^4$ \percc\ and PDR models. A general 
prediction of PDR models is that substantial heating extends to a visual 
extinction, A$_{\rm v}$, of 10 magnitudes, or a column density of 
10$^{22}$ \percmsq. Given the average H$_2$ density, we have a good 
estimate of the {\it total} distance from the ridge to the PDR interface 
which will not be affected by clumping.  Using 3 10$^4$ \percc, the {\it 
total} distance is 3 10$^{17}$ cm or 0.1 pc. If H$_2$ densities in 
the CO emitting region are larger, this is an {\it upper} limit to the 
line-of-sight distance. Since the Trapezium is 0.25 pc in front the of 
the PDR interface/Main Ionization Front (see O'Dell 2001), the total 
distance from $\theta ^1$C to the Orion ridge must be $\le$0.35 pc, but 
in no case less than 0.27 pc. Rodr{\i}guez-
Franco, A., Mart{\i}n-Pintado, J. \& Fuente, A. (1998) also favor a 
small line-of-sight distance between the Orion ridge and PDR on the 
basis of HC$_3$N and CN kinematics.

\subsection{The Orion KL outflow}

This source is prominent in our images because of the energetic CO 
outflow. Most likely, the region driving the energetic CO outflow is a 
heavily obscured  compact radio continuum source (Churchwell \etal 1987; 
Garay, Moran \& Reid 1987 (GMR)), which coincides with the SiO maser 
center (Menten \& Reid 1995 (MR)). In the IAU classification, this 
source is `[GMR] B' or `[MR95c] I'; the most commonly used name is 
source `I'. We show an image of the CO $J=7-6$ integrated intensities 
for a typical range of red and blue shifted velocities in Fig.~4. The 
position and overall distribution of the emission is very similar to 
that found in the $J=2-1$ line emission maps of RMW. As found for lower 
$J$ CO lines, the line connecting the blue and red shifted maxima passes 
10$''$ north of (0$''$, 0$''$) in Fig.~2, the position of source `I'. A 
comparison of the $J=2-1$ emission with the $J=7-6$ data shows that the 
$7-6$ emission has more structure. The critical density needed to 
populate the $J=7$ level is $\sim 10^6$ \percc,  43 times larger than 
for the $J=2$ level. Thus we conclude that these differences are caused 
by line excitation effects. 

In Table 1, we list integrated CO intensities for selected velocity 
intervals (Col.~3 and 4) and also ratios of integrated line intensities 
of the $J=7-6$ and $J=4-3$ lines to the $J=2-1$ line (Col.~5 and 6). We 
have chosen the same velocity intervals as those used by RMW to easily 
compare our data with their $J=2-1$ line results. The average of the 
ratios of the sub-mm CO lines to the $J=2-1$ line is given in Table 1.   
Trying a number of different choices of linear or parabolic baselines, 
we find that the RMS difference in our ratios is $\sim$15\%. 

The line ratios are very different from LTE ratios for a very warm 
molecular gas. Given the physical conditions in Orion, Large Velocity 
Gradient (LVG) models are one approach to determine average densities. 
We have taken the kinetic temperature in the outflow to be 150 K, and 
chosen a gradient of 1200 \kmsns/pc. The LVG model for an H$_2$ density 
of 10$^5$ \percc gives ratios of  $J=4-3$/$J=2-1$=2.6 and $J=7-6$/$J=4-
3$=2.3. An alternative is a PDR model. Here a number of additional 
measurements and assumptions are needed to estimate the H$_2$ density. 
The plane-parallel PDR Model A of KSSS describes a region irradiated on 
both sides. From their Fig.~7(a) for an H$_2$ density of 10$^7$ \percc, 
the predicted ratios are $J=4-3$/$J=2-1$=1.6 and $J=7-6$/$J=2-1$=2.1. 
The agreement of this prediction with our data is worse than for the LVG 
model, although the errors are large. On the basis of the average 
values, we conclude that the LVG model is a better description of the 
outflow. 

There is a significant difference in the source sizes for red (43$''$) 
and blue (34$''$) shifted gas, so we have obtained line ratios by 
spatially integrating intensities  over velocity slices. Also, from 
Fig.~4, the outflow centers are significantly offset from the (0$''$, 
0$''$) position. Thus the data collected in Table 6 of Schulz et 
al.~(1995), based on peak temperatures for (0$''$, 0$''$) alone, are 
less accurate. 

\subsection{Orion~S}

Compared to Orion KL, this is a prominent source in sub-mm dust emission 
(MZW), a compact emission region in NH$_3$ (Batrla \etal 1983), but is 
less prominent in CO emission, and shows only a few H$_2$O masers (Gaume 
\etal 1998). Gaume \etal (1998) detected no near-IR sources at the 
center of the H$_2$O masers, so this source is very deeply embedded.  
Orion~S is hot (\TKIN$\geq$300 K) and shows intense [O~I] and [C~II] 
emission (Herrmann \etal 1997). There is a low intensity, compact CO 
outflow. The relative positions of the red and blue shifted maxima are 
similar to Orion KL (see color plate). From studies of a number of 
molecular species, McMullin, Mundy \& Blake (1993) concluded that the 
chemistry was consistent with a young region where shock chemistry 
played the most important role. 

In Fig.~1(c) and (d) we show CO $J=4-3$ spectra of the blue- and red-
shifted line wings. The outflow spectra are strikingly similar to those 
in Fig.~2 of RMW. Because of the low line intensities, we have not 
mapped the outflow regions, but have taken longer integrations at the 
blue and red shifted maxima. We list the integrated intensities for the 
two maxima in Table~2. The FWHP beams used to take the $J=2-1$ and $J=4-
3$ spectra have similar sizes, so we have formed ratios without 
corrections for beam or source sizes. From the RMW data, we find that 
the outflow FWHP sizes are 23$''$ for the red shifted gas and 27$''$ for 
the blue shifted CO. The ratios for the blue shifted gas are 
significantly larger than for the red shifted gas. The LVG model for an 
H$_2$ density of 7 10$^3$ \percc\ gives a $J=4-3$/$J=2-1$ ratio of 0.8, 
while 5 10$^4$ \percc\  gives a ratio of 1.4. The PDR model A of KSSS 
(Fig.~7(a)) predicts a ratio of $\sim$1 for an H$_2$ density of 10$^5$ 
\percc. The $J=7-6$ line data had only very short integration times at 
each position, so were too noisy to allow a detection of the outflow. 
The difference between the line ratios for Orion KL and Orion~S may 
indicate that PDR conditions play a larger role in Orion~S, while the more
compact size of Orion~S is consistent with a younger source. 

In Fig.~5, from high angular resolution data, we show the maxima of 
different species; except for the H$_2$O masers, the emission centers 
are extended by $>$10$''$. Johnston \etal (1983) found 4 K absorption 
lines of the 6~cm line of H$_2$CO over 50$''$ toward Orion~S, but no 
compact continuum source. The 6~cm line of H$_2$CO usually has T$_{\rm 
ex} <$2.7 K, but the deeper absorption found toward Orion~S requires a 
discrete background source. The observation can be explained if some 
free-free emission, perhaps from Orion A, were behind the region 
containing the H$_2$CO. If the electron density in the ionized gas were 
10$^4$ \percc, the line-of-sight depth behind the H$_2$CO absorption 
region would be $\sim$10$^{-3}$ pc. The ionization fronts bordering such 
a large neutral region inside the Orion A \HII region would have been 
seen in the VLA and HST data of O'Dell \& Yusef-Zadeh (2000). Thus there 
is contradiction 
which cannot be resolved at this time. The submm CO and FIR results are 
most easily explained by placing Orion~S very close to the PDR 
interface, where Orion~S is exposed to a high UV radiation field. This 
accounts for the warm dust, high \TKIN, and a high abundance of atomic 
oxygen and ionized carbon, but less molecular emission. The offsets of 
the atomic and molecular maxima are consistent with the major source of 
ionization arising from the Trapezium. Since Orion~S is close to the 
Orion ridge, the emission is confused and mass estimates are uncertain. 
Muders \& Schmid-Burgk (1992) report the presence of rotation and 
estimate a mass of 7 M$_\odot$. From their submm dust map, MZW report 65 
M$_\odot$; this is very probably an overestimate because of confusion 
with the ridge. 

In Plate 1, we show the position of the jet feature by an arrow. For 
three positions in the jet we show spectra in Fig.~1 (e)-(g). The $J=4-
3$ to $J=2-1$ ratios of the integrated T$_{\rm MB}$ values are 1.3, 0.5 
and 0.7, respectively. The first two values refer to positions separated 
by only 15$''$. The jet feature is at the edge of our $J=7-6$ map. The 
nearest position is at an offset ($-40''$, $-150''$), where we find a 
$J=7-6$ to $J=4-3$ ratio of 0.6. Thus, there are large variations in the 
ratios. Applying an LVG analysis, we obtain an average H$_2$ density of 
$\sim$10$^4$ \percc. 

\subsection{The Orion Bar Feature}

The Bar feature is a neutral-ionized gas interface to the SW of Orion KL. 
This is one of the best studied PDR's and is used for testing PDR 
models.    From Fig.~1(h), the CO emission from the Bar has a radial 
velocity of $\sim$10.5 \kmsns, with a  FWHP of $\sim$3.3 km s$^{-1}$. 
There is unrelated CO emission at $\sim$4\kmsns. We show a set of radio 
line measurements in Fig.~6; our data have been produced by integrating 
intensities from 10 to 11 \kmsns.  The FWHP of the Orion Bar, measured 
in the direction of the ionization front (hereafter IF) is 67$''$. We 
estimate that the FWHP of the $J=6-5$ CO emission is $\sim 60''$ from 
the gray scale plot in Fig.~1 of Lis, Schilke \& Keene (1997; LSK). The 
$J=6-5$ \THCO\ data in Fig.~2 of LSK give a FWHP of 30$''$. In contrast, 
the CO $J=1-0$ data of Tauber \etal (1994; TTMG) gives a FWHP of 22$''$. 
We plot the $J=1-0$ data in Fig.~6(b). LSK give only relative 
coordinates, so we have not used these data in Fig.~6. To reconcile the 
different FWHP's we assume that the optical depth of the $J=7-6$ CO 
emission is $\sim$10 times larger than $J=1-0$ CO. The $J=1-0$ maximum 
of TTMG (1994) is $\sim$5$''$ further from the IF than our $J=7-6$ 
maximum, and nearly coincident with our $4-3$ peak. Our peak line 
intensity is T$_{\rm MB}$=145 K, as is the CO $J=6-5$ line of LSK, but 
the peak intensity of the $1-0$ line (TTMG) is 1.5 times smaller. 

KSSS have produced PDR models with a plane parallel geometry, illuminated 
from one side; these are appropriate for the Orion Bar. Their model 
(Fig.~12) gives the largest $7-6$/$4-3$ ratio, 1.4. Our observed ratio, 
$\sim$3.5, is more than twice the model prediction. The PDR model of 
KWHL may produce slightly higher ratios. It is possible that other 
mechanisms, such as shock waves, contribute to the heating of the gas in 
the Bar, or that clumping is a significant factor. In the following, we 
take the density from the PDR model of KWHL (Fig.~13 and 14), n(H$_2) 
\sim 10^5$ \percc. This value seems to be a lower limit, so the higher 
$J$ lines give H$_2$ densities which are significantly larger than 
$\sim5 \, 10^4$ \percc,  the value given by TTMG  (1994) and 
Hogerheijde, Jansen \& van Dishoeck (1995). 

The best measurement of the FWHP of hot CO emission from the Bar feature 
is 30$''$. If the geometry is cylindrical, with an effective diameter of 
30$''$ (=0.073 pc at 500 pc), and length of 100$''$ (=0.24 pc). Then the 
mass is 4.6 M$_\odot$. Corrections for clumping would lower this value. 
For a local density of 10$^5$ \percc, the column density of H$_2$ is 2.2 
10$^{22}$ \percmsq. The column density agrees with previous values for 
the Bar, but {\it requires a cylindrical geometry} not, as commonly 
thought, a sheet-like region with the long dimension parallel to the 
line-of-sight (see Fig.~13 of Hogerheijde \etal 1995). Walmsley \etal 
(2000) have also found evidence for a cylindrical geometry of the Bar 
from near-IR data. 

\subsection{Orion Cloud Mass Estimates}

To estimate the mass of H$_2$ over the 8 square minute region mapped in 
the $J=7-6$ CO line, we will assume that this is a PDR. The heating in 
PDR's is effective up to a H$_2$ column density of 10$^{22}$ \percmsq. 
Summing this column density over the region mapped, and including a 10\% 
helium contribution, we find 15 M$_\odot$ of warm molecular gas. Our 
estimate of this mass does {\it not} include any contribution from the 
Orion ridge or Bar. 

There have been estimates of the total mass of this region. From 
measurements of 1.3 mm dust emission, MZW quote a mass of $\sim$1.7 
10$^3$ M$_\odot$. This estimate depends on (uncertain)  dust 
temperatures and dust properties. We can estimate a more accurate mass 
between $\Delta \delta=1'$ and $\Delta \delta=-2'$ using the $J=2-1$ 
C$^{18}$O data of White \& Sandell (1995) and Eq.~14.115 of Rohlfs \& 
Wilson (1999). The value is $\sim$310 M$_\odot$. Since the lowest 
contour in the map is 20 K \kmsns, this {\it must} be a lower limit to 
the total molecular gas mass.  Goldsmith, Lis \& Bergin (1997) analyzed 
the three lowest transitions of \CEIO\ for the region between $\Delta 
\delta=6'$ and $\Delta \delta=-6'$. Scaling their result to the region 
of interest, we find a mass of 420 M$_\odot$. Our value and that of 
Goldsmith \etal (1997) probably represent the range of molecular mass in 
denser gas. Then $\sim$4-5\% of the total molecular mass is in the form 
of dense warm molecular gas. As pointed out by a number of authors (see, 
\eg Wilson \etal 1997) the mass of ionized gas in Orion A is $\ll$10 
M$_\odot$. Taking the estimates in Hillenbrand \& Hartmann (1999), we 
find 1200 M$_\odot$ of stars within 1.2$'$ of $\theta ^1 $C.  Thus, the 
mass in stars dominates. 

 \section{Summary}

We have mapped an 8 square arcmin region in the $J=7-6$ and $J=4-3$ lines 
of CO. We summarize the properties of individual  CO emission features 
in Table 3. From our sub-mm CO data and published results, we find that: 

\noindent
1. The $J=7-6$ quiescent CO emission is present over the entire region 
mapped. The gas is quiescent with $\Delta$V$_{1/2}$=4 to 6 \kmsns. Our 
data, together with PDR model calculations of KWHL, show that the $J=4-
3$ and $J=7-6$ CO emission is consistent with an H$_2$ density of $10^5$ 
\percc. The neutral-ionized gas interface has a depth of $\sim$3 
10$^{17}$ cm. The mass of warm molecular gas in the \HII region--
molecular cloud interface which we mapped in the $J=7-6$ line of CO is 
15 M$_\odot$. 

\noindent
2. The warm CO emission peaks west of the position of the Trapezium star, 
$\theta ^1$ C. There is a difference in the positions of the CO, with 
$J=7-6$ CO 33$''$ peaking west of the position of $\theta ^1$ C. The 
$J=2-1$ CO is 5$''$ west of the $J=7-6$ maximum, and \THCO\ 9$''$ west 
of the $J=7-6$ peak. From the differences in peak position, we conclude 
that the CO $J=7-6$ emission toward the Orion ridge is partly heated by 
Trapezium stars. Embedded stars may provide the rest of the heating. 

\noindent
3. From a PDR analysis of CO and atomic fine structure [C~II] and [O~I] 
line data, the {\it total} distance of Orion ridge from the \HII region-
molecular cloud interface is 3 10$^{17}$ cm. Since the projected 
distance is 2.8 10$^{17}$ cm, the line-of-sight distance is 10$^{17}$ 
cm. The CO emission arises from clumps with a volume filling factor 0.3. 
From the filling factor and maximum Planck temperature of 204 K, the 
maximum value of \TKIN\ is 680 K. If H$_2$ densities in the CO emitting 
region are $> 10^5$ \percc, the volume filling factor and line-of-sight 
distance will be smaller and  \TKIN\ will rise. 

\noindent
4. The ratios of $J=7-6$ to $J=4-3$ and $J=4-3$ to $J=2-1$ in the Orion
KL outflow are $\sim$2. As with the $J=2-1$ CO, the line connecting the
largest peaks in the outflow is 10$''$ north of the position of source
`I', which is thought to be the driving source (MR). From an LVG analysis,
the H$_2$ density in the Orion KL outflow is $\sim$10$^5$ \percc.

\noindent
5. The Orion~S region has high \TKIN\ and 
large abundance of atomic species are most simply explained by assuming 
that this region is younger than Orion~KL and very close to the PDR 
interface at the back face of the \HII region Orion A. 

\noindent
6. The ratio of the $J=4-3$ to $J=2-1$ lines in the Orion~S outflow is 
significantly larger for blue shifted CO. The ratios are lower than the 
ratios found for Orion~KL. In an LVG model, the H$_2$ densities are 7 
10$^3$ \percc\ for the red shifted CO and 5 10$^4$ \percc\ for the blue 
shifted CO. From a PDR analysis, the H$_2$ density would be 10$^5$ 
\percc. 

\noindent
7. For the highly collimated jet-like feature extending SW of Orion~S, we 
have measured $J=4-3$ to $J=2-1$ ratios at 3 positions. The average 
value is unity. An LVG analysis gives an H$_2$ density of $\sim$10$^4$ 
\percc. 
\noindent
8. Our $J=7-6$ to  $4-3$ line ratio for the Orion Bar is $\sim 3.5$,
larger than the highest value predicted by PDR models. We take the H$_2$ density
for this region to be $\geq$10$^5$ \percc. From this density and measured sizes, the mass is 4.5 M$_\odot$
Given this density, the geometry must be cylindrical, not a sheet-like geometry,  to match
the generally accepted column density of $\sim$10$^{22}$ \percmsq.

\noindent
9. From an analysis of C$^{18}$O $J=2-1$ line emission data, the total mass of gas in the region mapped is between 310 and 420
M$_\odot$. Based on PDR models, the mass of warm molecular gas in the PDR interface is
15 M$_\odot$, while the mass in ionized gas is $\ll$10 M$_\odot$, and
mass of stars in this region is 1200 M$_\odot$.

\noindent
{\bf Acknowledgements} We thank the Sub-mm Array Receiver Group at the
Harvard-Smithsonian Center for Astrophysics for providing the HEB
receiver. We also thank the SMTO staff, especially M. Dumke, for help with the observations. R. Mauersberger and A. Rodr{\i}guez-Franco
provided data in a digital form. W. Fusshoeller helped to prepare
some of the figures. An anonymous referee and Prof C. R. O'Dell helped to improve the text.

\begin{deluxetable} {cccccc}
\tiny
\tablecaption{High Velocity Line Wing Emission in Orion~KL }
\tablehead{\colhead{(1)}&\colhead{(2)}
&\colhead{(3)} &\colhead{(4)} &\colhead{(5)} &\colhead{(6)}\\
\colhead{Velocity }&\colhead{$J=2-1$}
&\colhead{$J=4-3$}&\colhead{$J=7-6$}&\colhead{Ratio\tablenotemark{(1)} }  &\colhead{Ratio\tablenotemark{(2)} }\\
\colhead{Range}  &\colhead{Transition } &\colhead{Transition }& \colhead{Transition}  &\colhead{of} &\colhead{of} \\
& \colhead{integrated}& \colhead{integrated} & \colhead{integrated}& \colhead{$J=4-3$}  & \colhead{$J=7-6$} \\
&intensity\tablenotemark{(3)}\  & \colhead{intensity}& \colhead{intensity}&\colhead{to} &\colhead{to} \\
& & & &  $J=2-1$ & $J=2-1$ \\
\colhead{(km s$^{-1}$)}  &\colhead{(K $\cdot$ \kms $'' \, ^2$)}   &\colhead{(K $\cdot$ \kms $'' \, ^2$)}   &\colhead{(K $\cdot$ \kms $'' \, ^2$)}   & & \\}
\startdata
 +35 to +55& 6.3   10$^5$  &  1.1 10$^6$      &  2.2   10$^6$ & 1.8      & 3.5\\
 $-$30 to $-$50& 3.2   10$^5$  &  6.8   10$^5$    &  9.9   10$^5$ & 2.1      & 3.1\\
 +55 to +75& 4.3   10$^5$  &  3.6   10$^5$    &  7.4   10$^5$ &0.8       & 1.7\\
 $-$50 to $-$70& 1.2  10$^5$   &  2.8   10$^5$    &  3.5   10$^5$ &2.3       & 2.9\\
 +75 to +95& 6.6   10$^4$  &  9.0   10$^4$    &  6.1   10$^4$ &1.4       & 0.9\\
 $-$70 to $-$90& 4.4   10$^4$  &  9.9   10$^4$    &  5.8   10$^4$ &2.2       & 1.3\\
 +95 to +115&1.6   10$^4$  &$\sim$8.3   10$^4$&   ---   &$\sim$5.2 & ---\\
 $-$90 to $-$110&1.6   10$^4$  &$\sim$3.0   10$^4$&   ---   &$\sim$2.1 & ---\\
Average\tablenotemark{(4)} & ---   & --- &   &2.2$\pm$1.3 & 2.2$\pm$1.1\\
\enddata
\tablenotetext{(1)}{ From data in col. 3 and 2.}
\tablenotetext{(2)}{ From data in col. 4 and 2.}
\tablenotetext{(3)}{ Data from RMW}
\tablenotetext{(4)}{ Unweighted average. }
\end{deluxetable}

\begin{deluxetable} {cccc}
\tiny
\tablecaption{High Velocity Line Wing Emission in Orion~S }
\tablehead{\colhead{(1)}&\colhead{(2)}
&\colhead{(3)} &\colhead{(4)} \\
\colhead{Velocity }&\colhead{$J=2-1$}
&\colhead{$J=4-3$}&\colhead{Ratio\tablenotemark{(3)} } \\
\colhead{Range}  &\colhead{Transition } &\colhead{Transition } &\colhead{of} \\
& \colhead{integrated} & \colhead{integrated}& \colhead{$J=4-3$}   \\
&\colhead{intensity\tablenotemark{(1)}}\ & \colhead{intensity\tablenotemark{(2)}} &\colhead{to} \\
&  & &  $J=2-1$  \\
\colhead{(km s$^{-1}$)}  &\colhead{(K $\cdot$ \kms )}   &\colhead{(K $\cdot$ \kms )}      & \\}
\startdata
 +30 to +50    & 14.5 &  12.0  & 0.8\\
 $-$50 to $-$20& 15.0 &  14.8  & 1.0\\
 +50 to +70    & 19.4 &  17.2  & 0.9\\
 $-$80 to $-$50& 7.1  &  16.2  & 2.3\\
\enddata
\tablenotetext{(1)}{ Estimated from data of RMW}
\tablenotetext{(2)}{ From peak of red and blue shifted outflow profiles (Fig.~1c and d.)}
\tablenotetext{(3)}{ From data in col. 3 and 2.}
\end{deluxetable}

\begin{deluxetable} {lcccc}
\tiny
\tablecaption{Summary of CO Line Results}
\tablehead{\colhead{(1)}&\colhead{(2)}
&\colhead{(3)} &\colhead{(4)} &\colhead{(5)} \\
\colhead{Feature} & \multicolumn{2}{c}{CO $J=7-6$} &      \colhead{Line }  & \colhead{H$_2$} \\
        & \colhead{Maximum}      & \colhead{$\Delta$v$_{1/2}$} & \colhead{ Ratio\tablenotemark{(1)}  }        & \colhead{density} \\
        & \colhead{T$_{\rm{mb}}$} &            &   &    \colhead{n(H$_2$)}   \\
        &               &                      &   &          \\
        &               &                      &   &           \\
        & \colhead{(K)}           & \colhead{(\kms)}               &                  & \colhead{(\percc)}\\ }
\startdata
Quiescent warm gas & 150     & 4--7            & $\sim1.2$        & $\geq10^5$\\
KL outflow         &  ---    &                & $\sim2$           & $10^5$   \\
Orion~S outflow    &  ---    &                & 0.8-1.4\tablenotemark{(2)}           & $<10^4$   \\
Orion~S jet        &  ---    &                & $\sim1$\tablenotemark{(2)}          &$\sim10^4$\\
Orion Bar          & 140     & 3.3            & $\sim3.5$          & $>10^6$\\
\enddata
\tablenotetext{(1)}{ Unless otherwise noted, ratio of $J=7-6$ to $J=4-3$}
\tablenotetext{(2)}{ Ratio of $J=4-3$ to $J=2-1$ lines}
\end{deluxetable}

\clearpage

 \begin {figure}
 \figurenum{1}
 \caption{{} CO line spectra taken toward various positions. The 
 intensity scale is T$_{\rm A}^*$. The offsets (upper right) are with 
 respect to $\alpha$=05$^{\rm h}$ 32$^{\rm m}$ 47$^{\rm s}$, $\delta=-
 05^{\rm o}$ 24$'$ 23$''$ (epoch 1950.0). {\bf (a)} The 
$J=4-3$ line taken
 toward the position of IRc2. {\bf (b)} The $J=7-6$ line taken toward 
 this position. {\bf (c)} The $J=4-3$ profile at the peak of the blue 
 shifted outflow from Orion~S (FWHP resolution 18$''$). {\bf (d)} The 
 corresponding red shifted outflow from Orion~S 
{\bf (e-g)} A series of line profiles from the jet feature, which 
extends to the SW of Orion~S. In each spectrum, $J=4-3$ emission is 
shown as a thin histogram, while the $J=2-1$ line (taken with the IRAM 
30-m telescope, beam 13$''$) is shown as a thicker smooth line. {\bf 
(h)} A $J=4-3$ emission line spectrum from the Bar feature. Superposed 
is the fit of 2 gaussians: The more intense line at 10.5 \kms is emitted 
from the Orion Bar while the weaker line arises from extended unrelated 
emission. } 
 \end{figure}

\begin {figure}
 \figurenum{2}
\caption{{}Velocity channel maps of the intensity of the the $J=7-6$ line 
of CO. The angular resolution is 13$''$; the radial velocity, \vlsr, is 
given in the upper left corner of each panel.  The units are integrated 
line intensity in K \kmsns, where the temperature is T$_{\rm A}^*$; the 
contour levels are 10 K \kms to 100 K \kms in steps of 10 K\kmsns. The 
region in the NE was not mapped. The zero point of the map coordinates 
is the one given in Fig.~1. The offsets are in arc sec. The prominent 
feature in the NW part of the map is the Orion KL outflow. The outflow 
covers a large velocity range, so is present in all the velocity 
channels shown. The (0$''$, 0$''$) position is marked by a `+'; the 
Orion~S outflow is marked by a `x'. The feature at \vlsr=9.43 and 10.85 
\kmsns, which extends from the Orion KL outflow to the Orion Bar 
Feature, is also seen in lower resolution maps of FIR fine structure 
lines of [O~I] and [C~II] (Herrmann \etal 1997).} 
  \end{figure}

\begin {figure}
 \figurenum{3}
\caption{{}Plots of gaussian fit parameters for the $J=7-6$, $4-3$ and 
$2-1$ lines of CO (from the HHT) as well as the $J=1-0$ line of \THCO\ 
(IRAM 30-m telescope) versus Right Ascension at the Declination of 
$\theta ^1$ C ($\Delta \delta=-53''$). The vertical line through all 
panels marks the R.~A. of the star $\theta ^1$ C Orionis. {\bf (a)} The 
peak temperatures, T$_{\rm MB}$, for the CO lines. The $J=7-6$ and $J=4-
3$ T$_{\rm MB}$ values were obtained by multiplying T$_{\rm A}^*$ values 
by 1.85 and 1.66, respectively. The conversion factor for other lines is 
1.2. Our maximum in the $J=7-6$ CO line is at $\Delta \alpha=-33''$ 
while the $J=2-1$ maximum from 30-m (FWHP 10$''$) and HHT (FWHP 33$''$) 
is at $\Delta \alpha=-38''$. The difference between the position of the 
$J=2-1$ maximum (representing cooler molecular gas) and $J=7-6$ maximum 
(warm gas) is significant. The \CEIO\ and \THCO\ peak at $\Delta 
\alpha=-42''$, so the column density of CO is west of the warmest CO 
peak. {\bf (b)} A plot of the FWHP line widths, $\Delta$v$_{1/2}$, as a 
function of offset in $\alpha$. {\bf (c)} A plot of \vlsr\ versus offset 
in $\alpha$. {\bf (d)} Atomic fine structure lines from Herrmann \etal 
(1997; FWHP resolution $\sim$1$'$) for the same Declination. The [O~I] 
line data is shown as solid and dashed lines; the [C~II] data is shown 
as a dash dotted line. }  \end{figure} 

 \begin {figure}
 \figurenum{4}
\caption{{} Plots of the integrated intensities for the Orion KL outflow 
in the CO $J=7-6$ line for two velocity intervals, {\bf (a)} the $-50$ 
to $-70$ \kms velocity range (contours 10, 20, 30, 40, 50, 60, 70, 80 K 
\kmsns) and {\bf (b)} the  55 to  75 \kms velocity range (contours 10, 
30, 50, 70 and 90  K \kmsns). The temperatures are T$_{\rm A}^*$; 
multiplying by 2.5 converts these contours to T$_{\rm MB}$. The star 
marks the (0$''$, 0$''$) position as in Fig.~1. This is the position of 
source `I', a 7 mm continuum and SiO maser source. `I' is considered to 
be driving the CO outflow (MR). Scanning effects cause the somewhat 
rectangular shape of the contours. } \end{figure} 

\begin {figure}
 \figurenum{5}
\caption{{} Features associated with Orion~S; offsets are relative to our 
zero point in Fig.~1. There is a compact, high velocity CO outflow (RMW) 
centered at ($\Delta \alpha, \, \Delta \delta)=(-16.5''$, $-85''$); the 
$J=4-3$ profiles taken at the maxima for the blue (`B') and red (`R') 
shifted CO are in Fig.~1(c) and (d). Our $4-3$ emission line peak is 
rather extended; our average peak position in CO $J=7-6$ is for 
\vlsr=2.3 to 12.28 \kms (see Fig.~2). We show the most intense H$_2$O 
maser emission center imaged by Gaume \etal (1998) with a 0.1$''$ beam. 
MZW reported the compact 1.3 mm emission region, OMC-1~FIR~4. Within the 
positional uncertainties of MZW, this is coincident with the H$_2$O 
position (Gaume \etal 1998). Mundy \etal (1986) found the CS maximum 
CS3. The H$_2$CO absorption toward Orion~S (Johnston \etal 1983, resolution 10$''$) has 
a FWHP of 50$''$ and a depth of 4 K.  }  \end{figure} 

 \begin {figure}
 \figurenum{6}
\caption{{} Spectral line emission from a number of species in the Orion 
Bar versus offset from the Ionization Front (IF).  {\bf (a)} Data taken 
along a line with Position Angle 135$^{\rm o}$, zero point 
$\alpha$=05$^{\rm h}$ 32$^{\rm m}$ 55.4$^{\rm s}$, $\delta=-05^{\rm o}$ 
26$'$ 50$''$ (1950.0). {\bf (b)}  Data taken along lines passing through 
$\alpha$=05$^{\rm h}$ 32$^{\rm m}$ 52.7$^{\rm s}$, $\delta=-05^{\rm o}$ 
26$'$ 50$''$ and  $\delta=-05^{\rm o}$ 27$'$ 00$''$ (1950.0), 
P.A.=135$^{\rm o}$. Our $J=7-6$ CO line data are shown as a thick solid 
line and our $4-3$ data are shown as a thick solid line passing through 
circles. Both results are in T$_{\rm A}^*$ units, integrated over a 
velocity range from 10 to 15\kmsns. The CO $1-0$ data are from TTMG 
(1994; resolution 7$''$). The CN data are from Simon et al.~(1997; 
resolution 14$''$).  The $J=5-4$ CS data are from der Werf et al.~(1996; 
resolution 8$''$), the $N=2-1$, $J=5/2-3/2$ CO$^+$ data from St\"orzer, 
Stutzki \& Sternberg (1995; resolution 12$''$) and the C~I data are from 
Tauber et al.~(1995; resolution 15$''$). {\bf (c)} Adapted from Wyrowski 
et al.~(1997); these results are averaged over the width of the Bar 
feature. The vibrationally excited \MOLH\ data, labelled H$_2^*$, were 
taken from van der Werf et al.~(1996). The position of the C91$\alpha$ 
carbon radio recombination line (resolution 11.7$''$ by 9.0$''$)  
represents the position of C$^+$.} \end{figure} 

\begin {plate}
 \platenum{1}
\caption{{}A color coded image of the intensity of the $J=7-6$ line of 
CO, integrated over the velocity range from $-150$ \kms to +150 \kmsns. 
The intensity scale is shown as a bar on the right side of the map. The 
angular resolution is 13$''$. The zero point of the map coordinates is 
$\alpha$=05$^{\rm h}$ 32$^{\rm m}$ 47$^{\rm s}$, $\delta=-05^{\rm o}$ 
24$'$ 23$''$ (1950.0). The maxima of the red and blue shifted CO 
emission in Orion KL and Orion~S are marked `R' and `B'. The four stars 
mark the positions of the Trapezium members. The lines with labels `(a)' 
and `(b)' in the SW are the paths in Fig.~6 (a) and (b). The second Bar 
peak to the NE was not measured in other species.} 
  \end{plate}

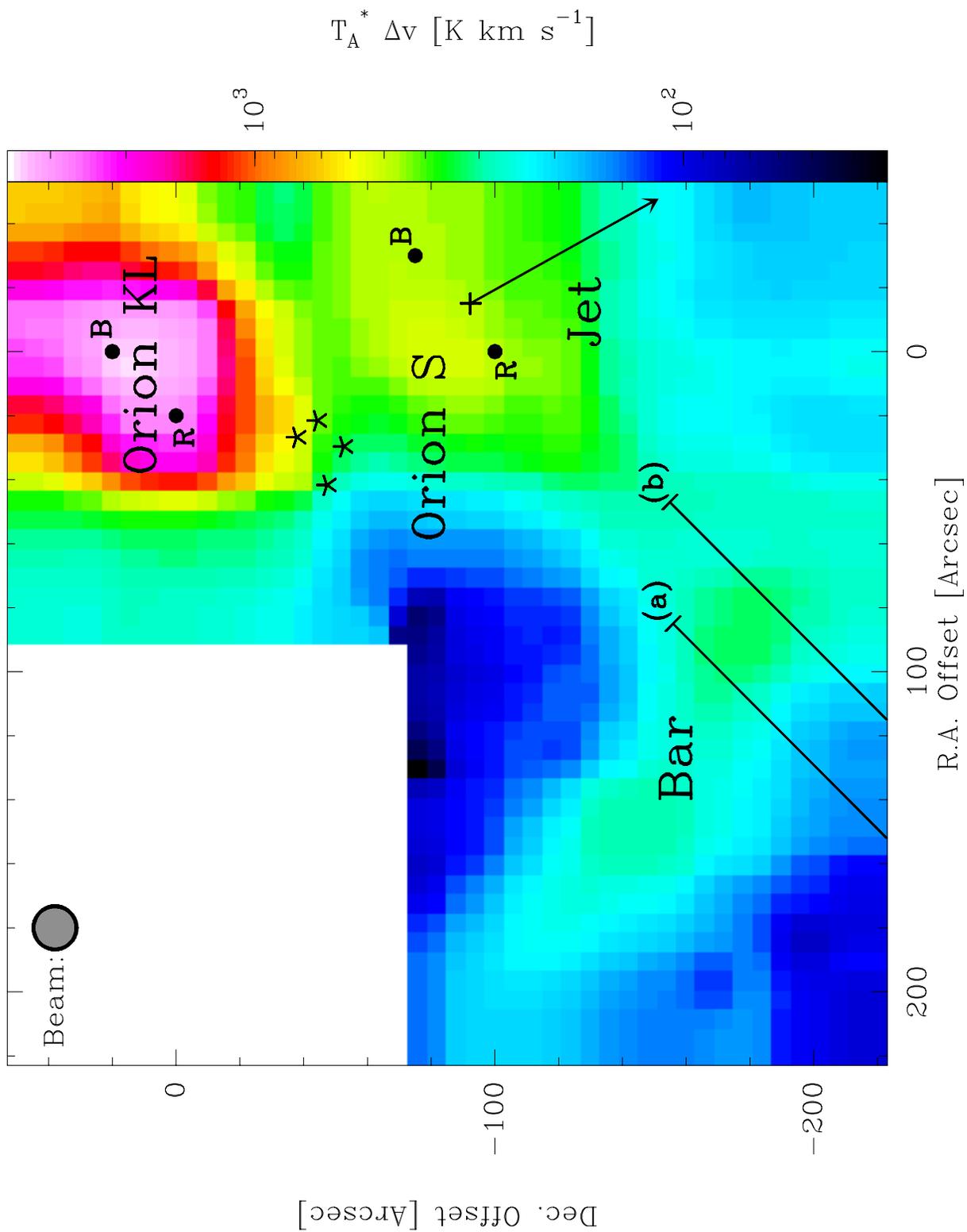
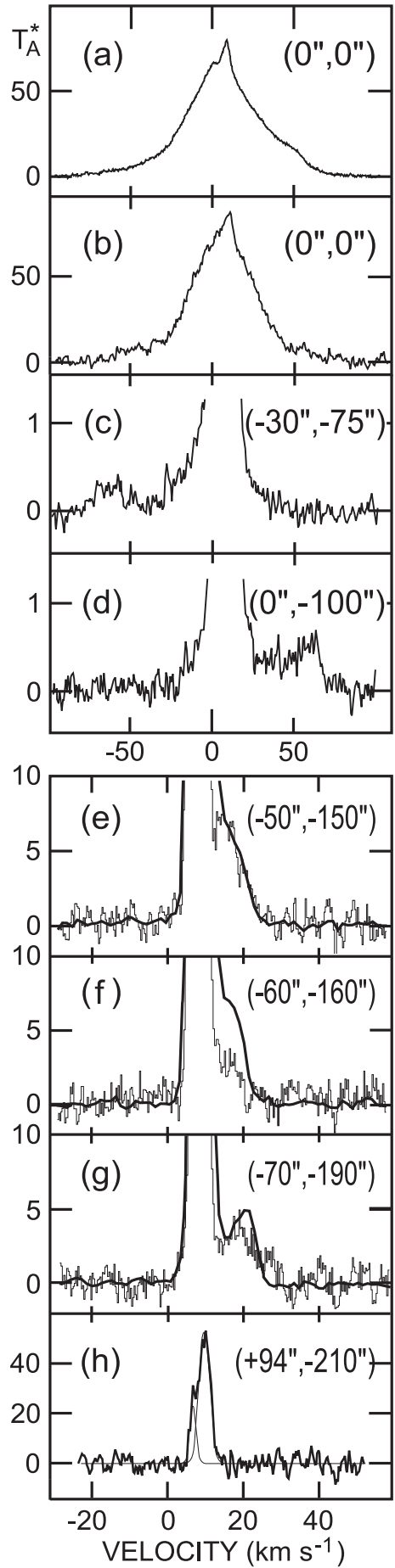
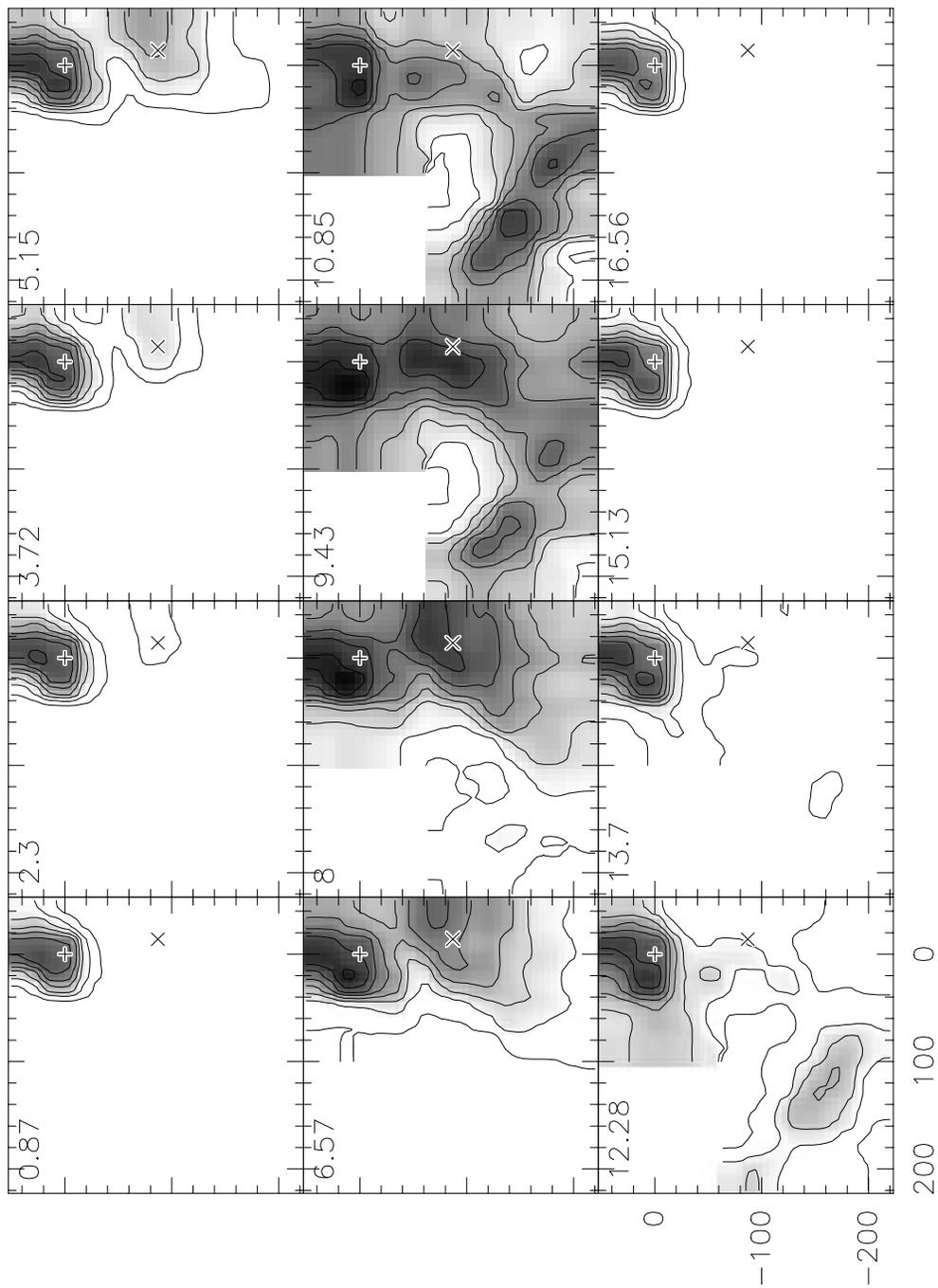
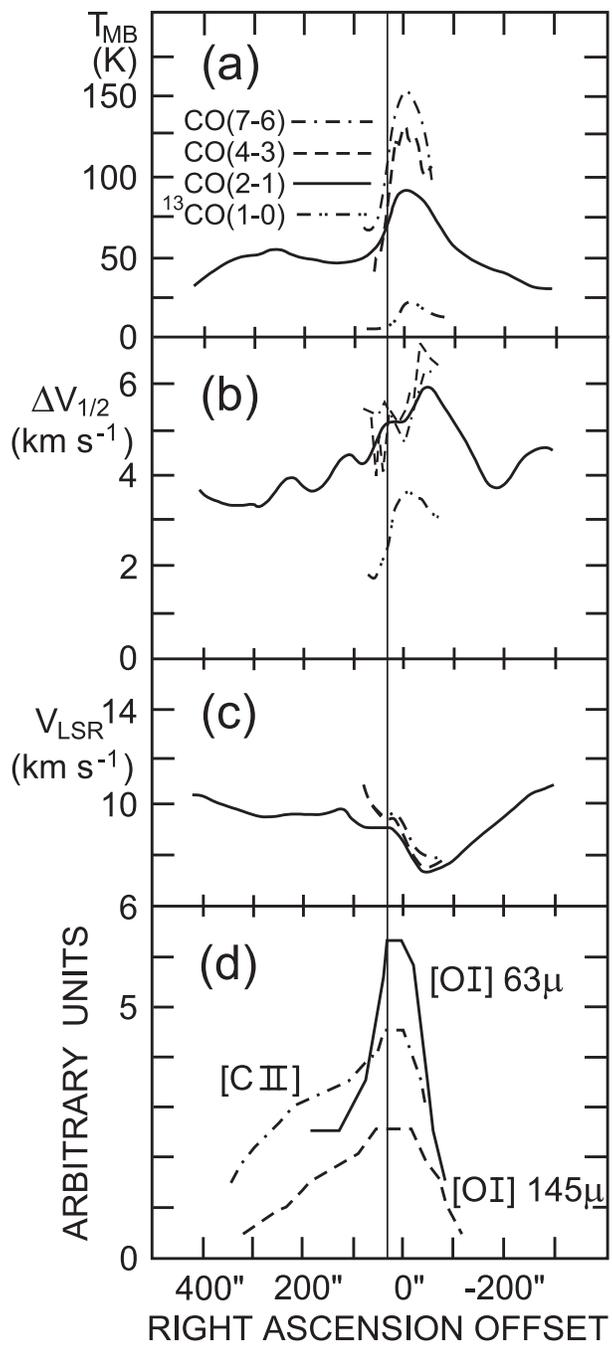
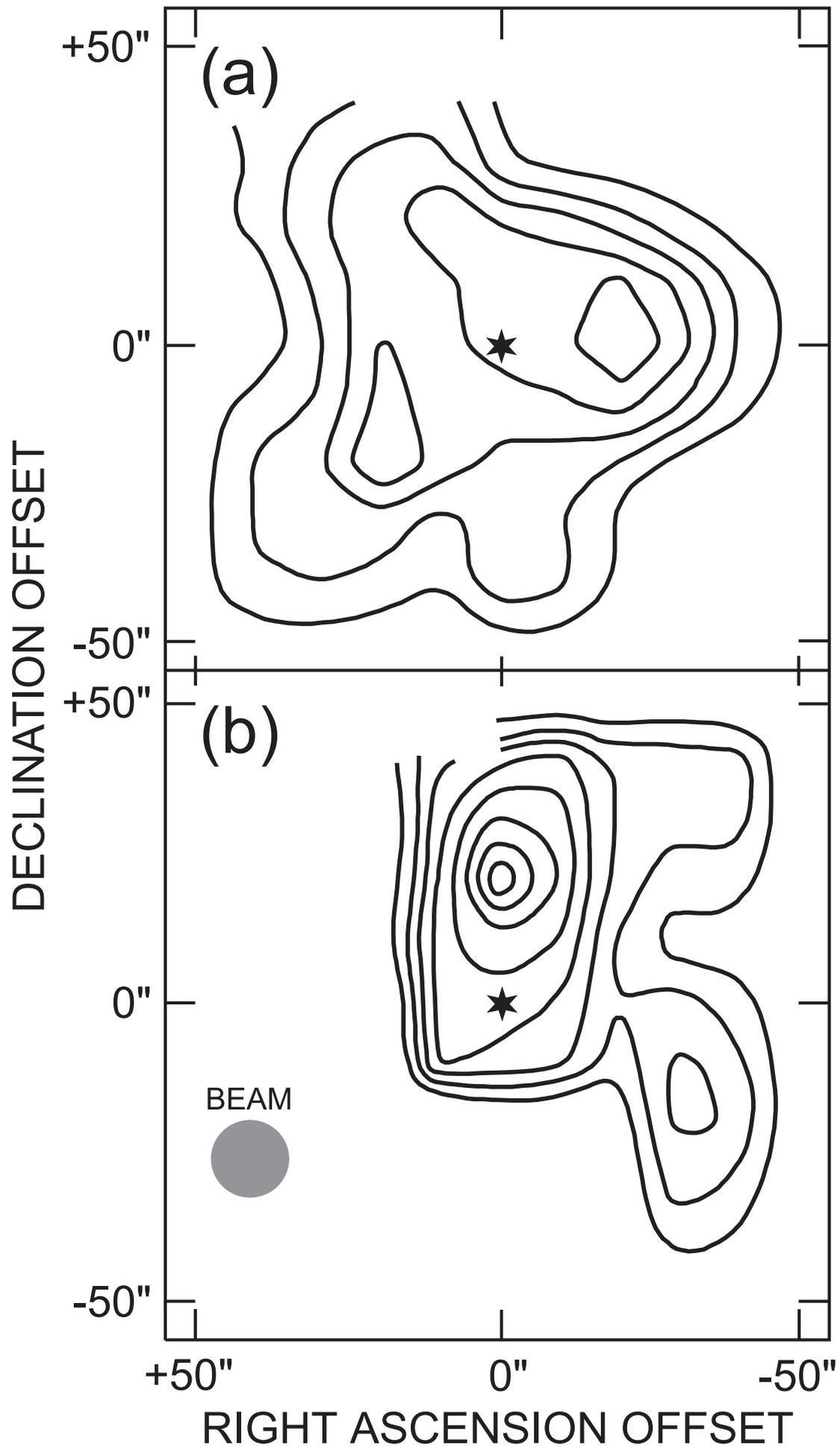
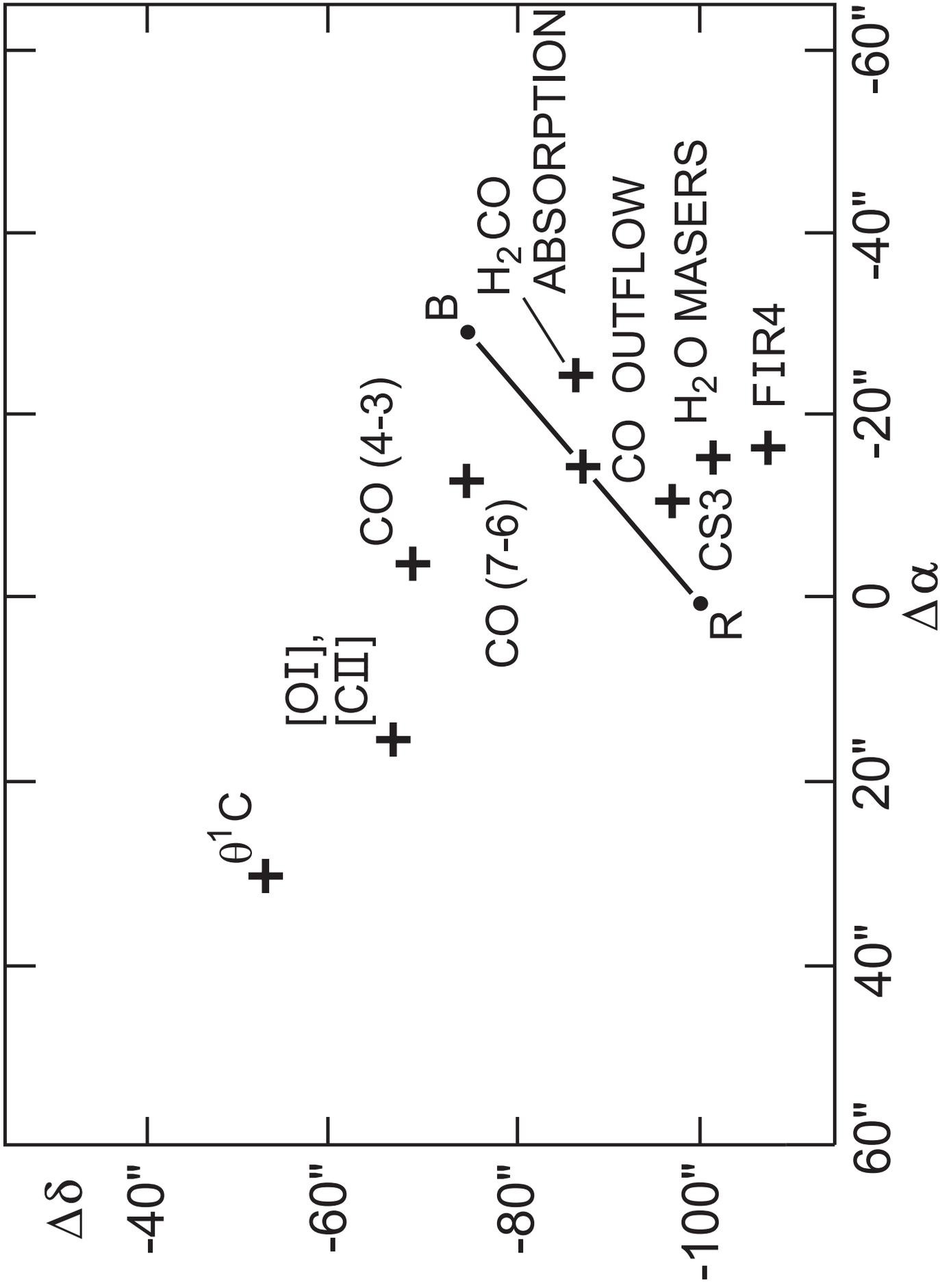
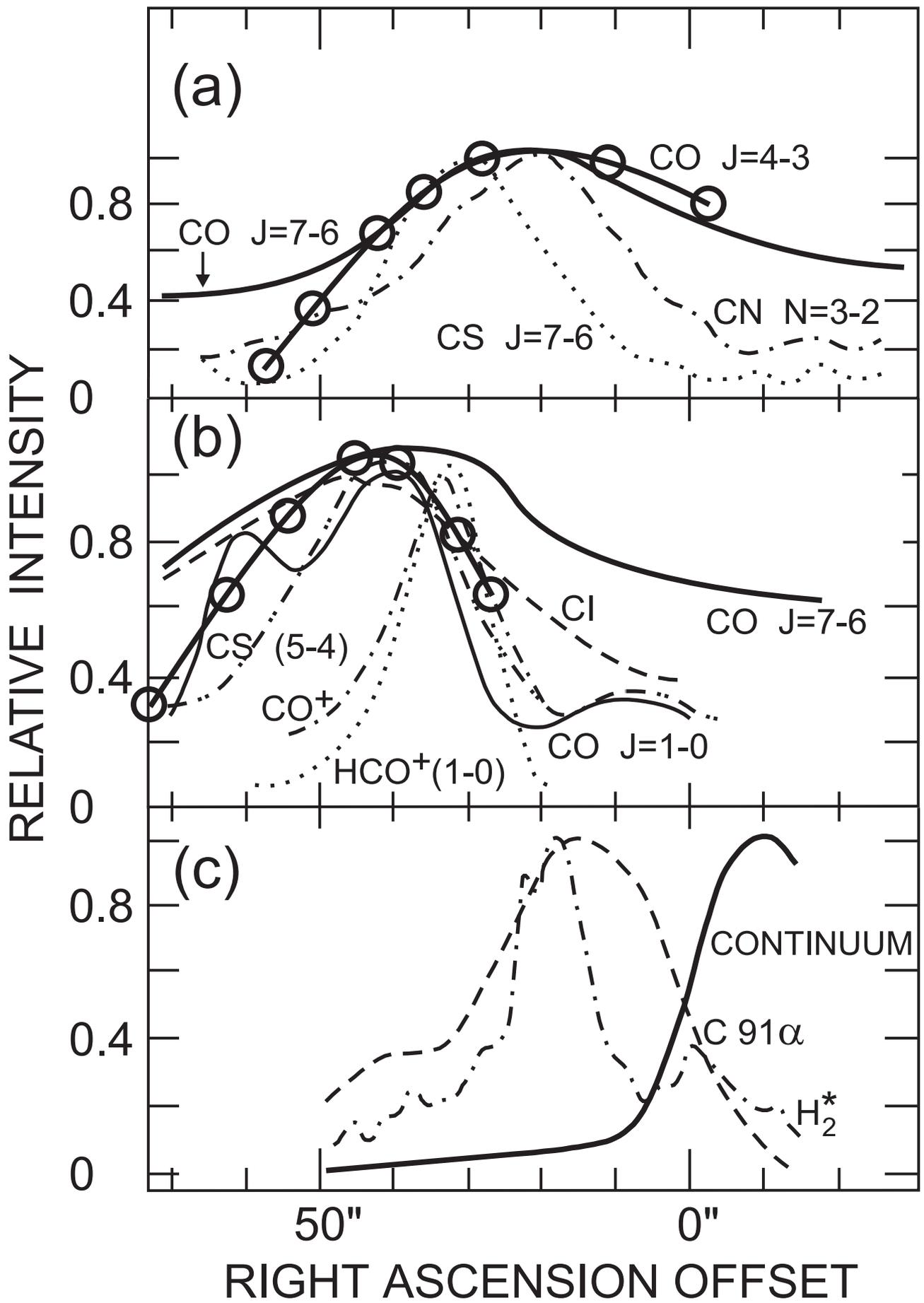
\end{document}